\documentclass[epjST]{svjour}
\usepackage{graphics}
\usepackage{epsfig}
\def\bfg #1{{\mbox{\boldmath $#1$}}}
\def\aphet{\bar p\,^3\rm{He}}

\begin{document}
\title{Spin dependence of the $\bar p d$ and $\bar p\,^3{\rm He}$ interactions}
\author{J.~Haidenbauer\inst{1} 
\and Yu.N.~Uzikov\inst{2}\fnmsep\thanks{\email{uzikov@jinr.ru}}  }
\institute{Institute for Advanced Simulation, Forschungszentrum J\"ulich, 
D-52425 J\"ulich, Germany 
\and Laboratory of Nuclear Problems, Joint Institute for Nuclear
Research, 141980 Dubna, Russia \& 
Department of Physics, Moscow State University, 119991 Moscow, Russia}
\abstract{
Elastic scattering of antiprotons on deuteron and $^3{\rm He}$ targets 
is studied within the Glauber-Sitenko theory. 
In case of $\bar p d$ scattering, the single- and 
double $\bar p N$ scattering mechanisms and the full spin dependence of the 
elementary $\bar p N$ scattering amplitudes are taken into account on the 
basis of an appropriately modified formalism developed for $pd$ scattering.
Differential cross sections and analyzing powers are calculated for 
antiproton beam energies between 50 and 300 MeV, using 
the $\bar N N$ model of the J\"ulich group as input.
Results for total polarized cross sections are obtained via the optical theorem.
The efficiency of the polarization buildup for antiprotons in a storage ring is
discussed.
} 
\maketitle
\section{Introduction}
\label{intro}
Scattering of antiprotons off polarized nuclei can be used to produce 
a beam of polarized antiprotons. Indeed, the PAX collaboration \cite{barone} intends 
to utilize elastic scattering of antiprotons off a polarized $^1$H target in 
rings \cite{frank05} as the basic source for antiproton polarization buildup. 
Analogous experiments performed for the proton case by the FILTEX collaboration \cite{FILTEX}
at 23 MeV and a recent COSY study where protons were scattering off a polarized 
hydrogen at 49 MeV \cite{frankpc} showed that a polarized beam can be 
achieved via this so-called spin-filtering effect. 
Whereas the spin dependence of the nucleon-nucleon ($NN$) interaction is very well known
at the considered energies, that allows one to calculate reliably \cite{MS} the
spin-filtering effect for protons,
there is practically no corresponding information for 
the antinucleon-nucleon ($\bar NN$) interaction. For this reason a test experiment 
for the spin-filtering effect in the antiproton-hydrogen interaction is planned 
at the AD ring at the CERN facility \cite{AD1,AD}.

In view of the unknown spin dependence of the $\bar pN$ interaction, the 
interaction of antiprotons with a polarized deuteron is also
of interest for the issue of the antiproton polarization buildup. 
This option was discussed in our paper  \cite{ujh2009} within
the single-scattering approximation.
The spin dependence of the elementary $\bar p N$ amplitudes was taken into account
only in collinear kinematics using the $\bar N N$ interaction model of the J\"ulich
group \cite{Hippchen,Mull1,Mull2,Haidenbauer2011} and total spin-dependent
cross sections were calculated for 
energies in the region 50--300 MeV using the generalized optical theorem.
A very similar analysis was performed by us for $\bar p\,^3{\rm He}$ elastic 
scattering and the corresponding total cross sections were calculated \cite{{ujhprm}}.
However, spin observables for elastic scattering of antiprotons on light nuclei
and shadowing effects (double scattering) in polarized total cross sections  
were not considered in those works. Spin observables are interesting quantities 
because they could be used for discrimination between existing models of the $\bar N N$ 
interaction once pertinent data become available \cite{AD1,AD}.
A calculation of such observables for the reaction $\bar p d\to \bar pd $, 
including double-scattering effects, was performed recently by us \cite{uzjh2012} 
and those results are reviewed here. 

\section{Method}
\label{sec1}
Our study is based on the formalism derived by Platonova and Kukulin \cite{pkuk} 
within the Glauber-Sitenko theory of multistep scattering \cite{GlauberFranco} and
applied to $pd$ elastic scattering -- appropriately modified by us for the 
$\bar pd\to \bar p d$ transition. 
The single (SS) and double scattering (DS) mechanisms are included and the $S$- and
$D$-components of the deuteron wave function and the full spin structure of the
elastic $\bar p N$ scattering amplitude are taken into account. 
We refer the reader to Ref.~\cite{uzjh2012} for details of the formalism. Here we
only provide the expression of the $\bar p N$ scattering matrix which we
need for the discussion lateron. It is given by
\begin{eqnarray}
 M_{\bar p N}&=&A_N+(C_N\bfg \sigma_1+C_N'\bfg \sigma_2)\cdot {\bf \hat n}
+B_N(\bfg \sigma_1\cdot {\bf \hat k})(\bfg \sigma_2\cdot {\bf \hat k})\nonumber \\ 
&+&(G_N-H_N)(\bfg \sigma_1\cdot {\bf \hat n})(\bfg \sigma_2\cdot {\bf \hat n})
+(G_N+H_N)(\bfg \sigma_1\cdot {\bf \hat q})(\bfg \sigma_2\cdot {\bf \hat q}) \ .
\label{pbarN}
\end{eqnarray}
In Eq.~(\ref{pbarN}), $\bfg \sigma_1$ ($\bfg \sigma_2$) is the Pauli matrix acting on the spin of the
$\bar pN$ states ($N=p,n$). The unit vectors are defined by 
${\bf \hat k}=({\bf k}_i+{\bf k}_f)/|{\bf k}_i+{\bf k}_f|$,
${\bf \hat q}=({\bf k}_i-{\bf k}_f)/|{\bf k}_i-{\bf k}_f|$,
and ${\bf \hat n}=[{\bf \hat k}\times {\bf \hat q}]$, where ${\bf k}_i$ (${\bf k}_f$) 
denotes the momentum of the incident (outgoing) antiproton.
The charge-exchange amplitude
$M_{\bar pp\leftrightarrow \bar nn}$ has the same spin structure as given in Eq.~(\ref{pbarN}).

The total $\bar p d$ and $\bar p^3{\rm He}$ cross sections are defined by 
 \cite{ujh2009}
\begin{equation}
\label{totalspin}
\sigma_{tot}=\sigma_0+\sigma_1{\bf P}_{\bar p}\cdot {\bf P}_T+
 \sigma_2 ({\bf P}_{\bar p}\cdot {\bf \hat  k}) ({\bf P}_T\cdot {\bf \hat k})+
\sigma_3 P_{zz}, 
\label{sigtot}
\end{equation}
where 
 ${\bf P}_{\bar p}$ ($ {\bf P}_T$) is the polarization vector of the antiproton (target),
 and $P_{zz}$ is the tensor polarization of the deuteron target ($OZ||{\bf \hat k}$)
(for the $^3{\rm He}$ target this term is absent).
The total cross sections
$\sigma_i$ ($i=0,1,2,3$) are calculated using the generalized optical theorem as described 
in Refs.~\cite{ujh2009,ujhprm}.


\begin{figure}[hbt]
\mbox{\epsfig{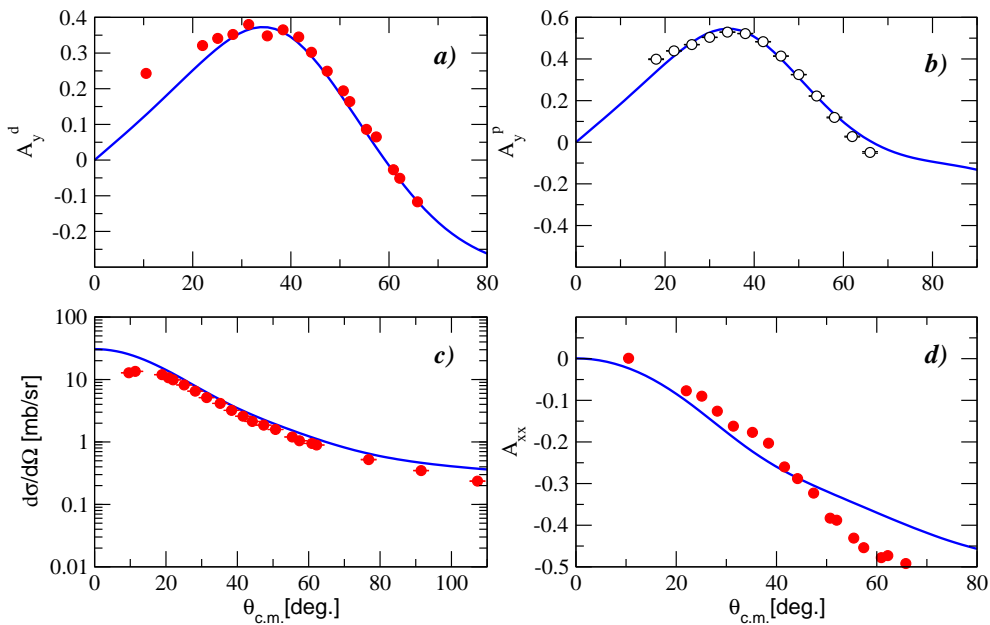}}
\caption{Analyzing powers  $A_y^d$ (a), $A_y^p$ (b),
$A_{xx}$ (d) and 
differential cross section (c) of elastic $ p d$
scattering at 135 MeV versus the c.m. scattering angle.
Results of our calculations based on the Glauber-Sitenko approach 
are compared with data from Refs.~\cite{riken} (filled circles) and \cite{IUFC}
(open circles).
The signs of $A_y^p$ and $A_y^d$ have been reversed in 
view of a different choice of the OY axis in Refs.~\cite{riken,IUFC} as 
compared to Ref.~\cite{pkuk} (see the definition of ${\bf \hat n}$ after Eq.~(\ref{pbarN})).
}
\label{pd135}
\end{figure}

\section{ Results and discussion}
\label{sec:2}
 
In Ref.~\cite{pkuk} the Glauber-Sitenko formalism was successfully applied for 
describing spin observables of $pd$ scattering at 250--1000 MeV,
taking into account the full spin structure of the $pN$ scattering amplitudes
and the $S$- and $D$-components of the deuteron wave function.
In order to test our own implementation of the formalism, we first tried to 
reproduce the numerical results of Ref.~\cite{pkuk}. As example we present 
here $pd$ results at 135 MeV, cf. Fig.~\ref{pd135}, where one can see that 
this approach allows to explain reasonably well the differential cross section, 
the vector analyzing powers, and to some extent also the tensor analyzing 
power $A_{xx}$ at this energy \cite{new}. 

\begin{figure}
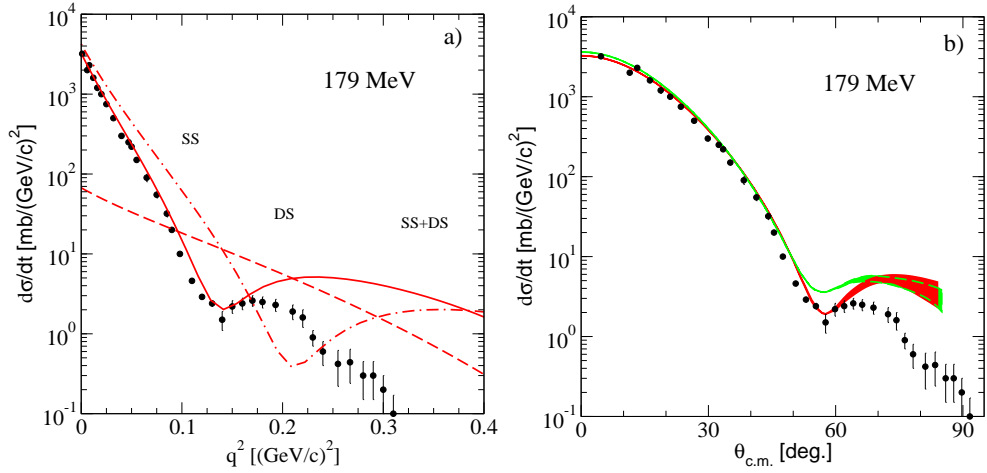

\vspace{0.5cm}
\mbox{\epsfig{figure=s179fig2.eps,height=0.295\textheight, clip=}
\epsfig{figure=s179fig2a.eps,width=0.300\textheight, clip=}
}
\caption{Differential cross section of elastic ${\bar p} d$
scattering at 179 MeV.
(a):
results based on single-scattering (dash-dotted line), 
double-scattering (dashed) and the full (solid) 
Glauber-Sitenko mechanisms are shown for the model D. 
(b): Results of the full calculation 
 are shown based on the $\bar NN$ models model A (green/grey) and 
 D (red/black). The bands represent the sensitivy to variations
 of the large-angle tail of the $\bar p N$ amplitudes as discussed
 in the text. 
 Data are taken from Ref. \cite{bruge}.}
\label{pd179q2}
\end{figure}

In the next step we use the modified formalism
to calculate observables for $\bar p d$ elastic scattering. 
Earlier studies of the antiproton elastic scattering \cite{laksmbiz81}
and also our own previous calculations \cite {ujh2009,ujhprm} were all 
done within the spinless approximation for the elementary ${\bar p N}$ 
amplitude $M_{\bar p N}$, i.e. keeping only $A_N$ from Eq.~(\ref{pbarN}), 
and restricted to using only the $S$-wave part of the wave function of the 
target nucleus.
In the present calculation we keep the full spin dependence of the $\bar pN$ 
amplitude (see Eq.~(\ref{pbarN}))
and employ two models developed by the J\"ulich group, namely 
A(BOX) introduced in Ref.~\cite{Hippchen} and D described in
 Refs.~\cite{Mull2,Haidenbauer2011}. 
 An exemplary result demonstrating the role of the single-scattering (SS) and 
 double-scattering (DS) mechanisms is shown in Fig.~\ref{pd179q2}a). 
 One can see that the SS mechanism alone fails to explain the forward peak.
 However, the coherent sum SS+DS describes it rather well. Obviously, 
 the DS mechanism, neglected in Ref.~\cite{ujh2009} in the calculation of the 
 spin-dependent total cross sections, has a sizable influence even in the 
 region of the forward peak.

 Considering the spin-dependent terms of the $\bar pN$ 
 amplitude, see Eq.~(\ref{pbarN}), one has to address the following issue: 
 In contrast to the spin-independent part $A_N$ ($N=p,n$), most of the other 
 terms that give rise to the spin dependence ($B_N$, $C_N$, $C_N^\prime$, $G_N$, 
 $H_N$) do not exhibit
 a well-pronounced diffractive behaviour for antiproton beam energies 50--200 MeV,
 i.e. they do not decrease rapidly with increasing center-of-mass (c.m.) 
 scattering angle $\theta_{c.m.}$. 
 As a consequence, the  differential ${\bar p} N$ 
 cross section has a minimum at scattering angle 
$\sim 100^\circ$ and a backward maximum \cite{uzjh2012}. 
 One should note, that the Glauber-Sitenko approach
 is not suitable for taking into account backward scattering in 
 the elementary hadron-nucleon collision, because its basis is the eikonal 
 approximation. 
Therefore, any sensitivity of the observables calculated within this approach
to the backward tail of the elementary $\bar pN$ amplitude is in contradiction 
with the assumptions of the Glauber-Sitenko theory and tells us that the 
corresponding calculations are no longer reliable. 
\begin{figure*}
\mbox{\epsfig{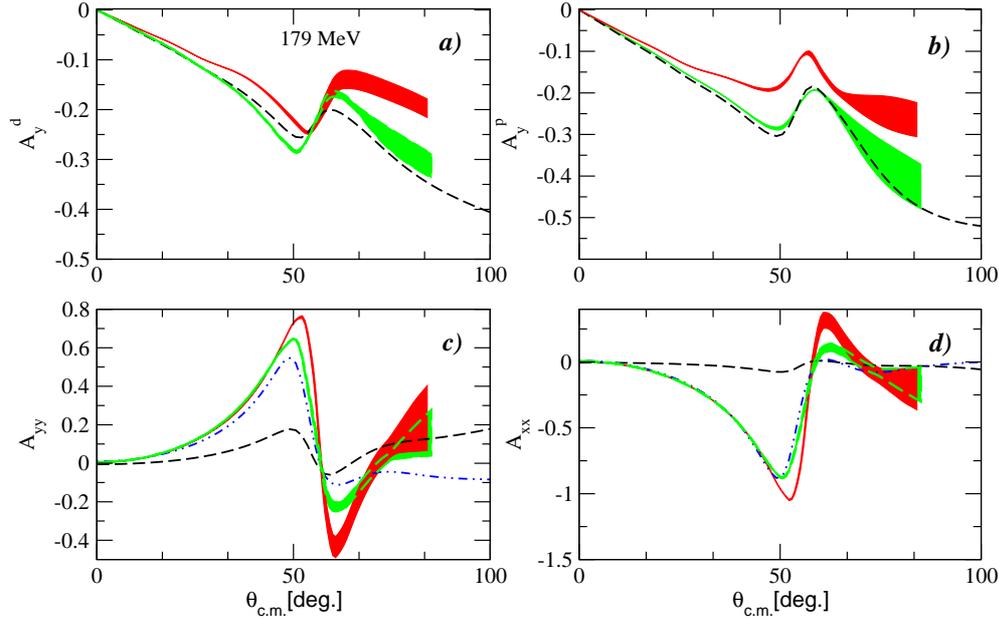}}
\caption{Spin observables of elastic ${\bar p}d$ scattering at 179 MeV 
 versus the c.m. scattering angle:
$A_y^d$ (a), $A_y^{\bar p}$ (b), $A_{yy}$ (c), and $A_{xx}$ (d).
 Results of our full calculation (including the SS+DS mechanisms)
 are shown based on the $\bar NN$ models model A (green/grey) and
 D (red/black). The bands represent the sensitivy to variations
 of the large-angle tail of the $\bar p N$ amplitudes as discussed
 in the text.  
The dashed line shows the result when the deuteorn 
$D$-wave is omitted, based on model A, while
the dash-double dotted line in (c) and (d) corresponds to 
exclusion of the spin-dependent terms in Eq.~(\ref{pbarN}).
 }
\label{ay179}
\end{figure*}
%
In the course of our investigation we studied this issue in detail 
by varying the employed elementary $\bar p N$ amplitudes in the 
backward-angle region and by examining the induced variations in 
the predictions for elastic $\bar p d$ scattering, see Ref.~\cite{uzjh2012}.  

The result of our analysis is summarized in Figs.~\ref{pd179q2}--\ref{ay179}.
The bands represent the sensitivity of the calculated $\bar p d$ observables
to variations of the backward tail of the elementary $\bar p N$ amplitudes. 
Thus, the widths of these bands is a sensible measure for estimating the 
angular region where the Glauber-Sitenko theory is able to provide solid 
results for a specific $\bar p d$ observable (vanishing width) and where 
it starts to fail (sizable width). 
\begin{figure}
\includegraphics[width=0.32\textwidth,angle=0]{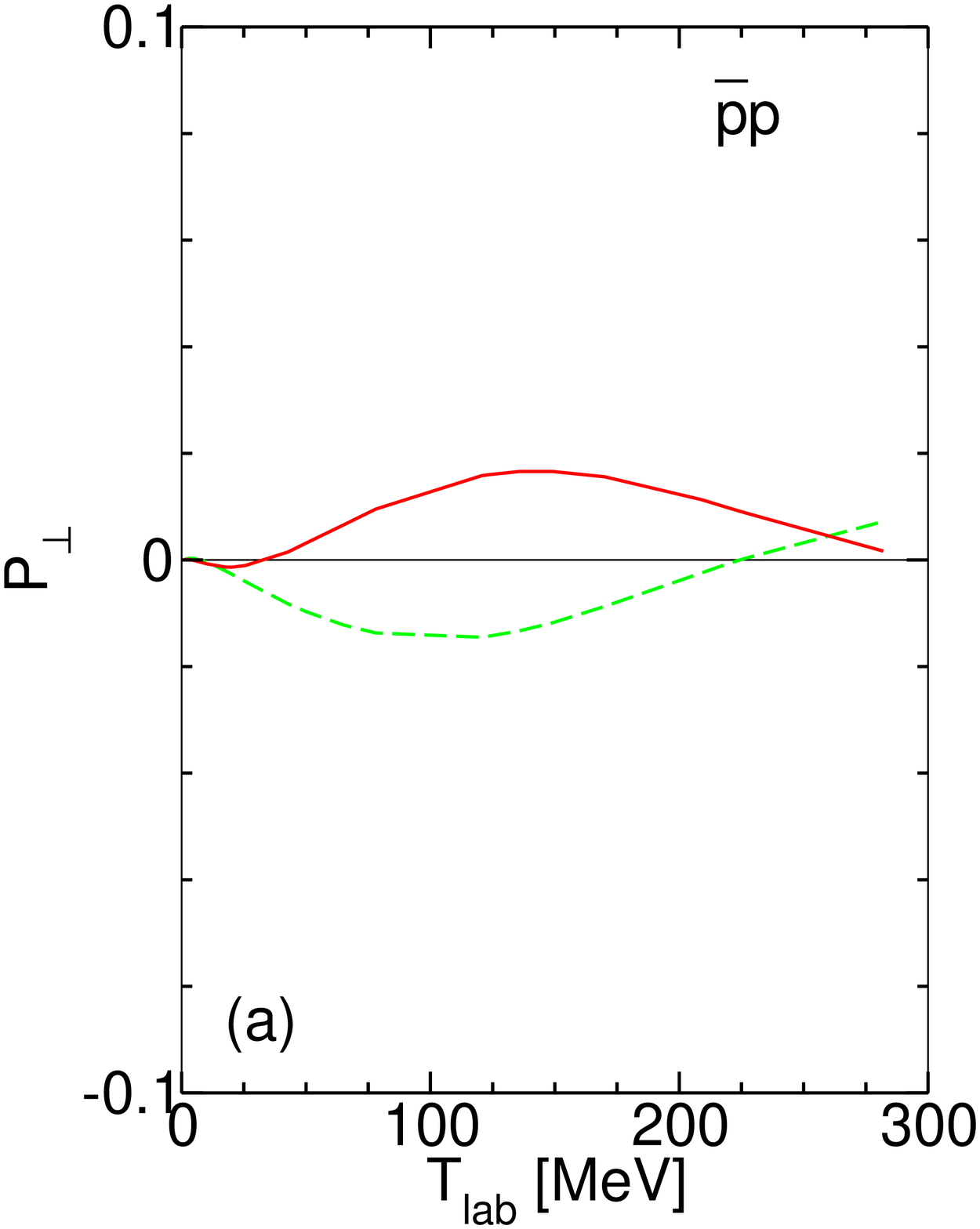}
\includegraphics[width=0.32\textwidth,angle=0]{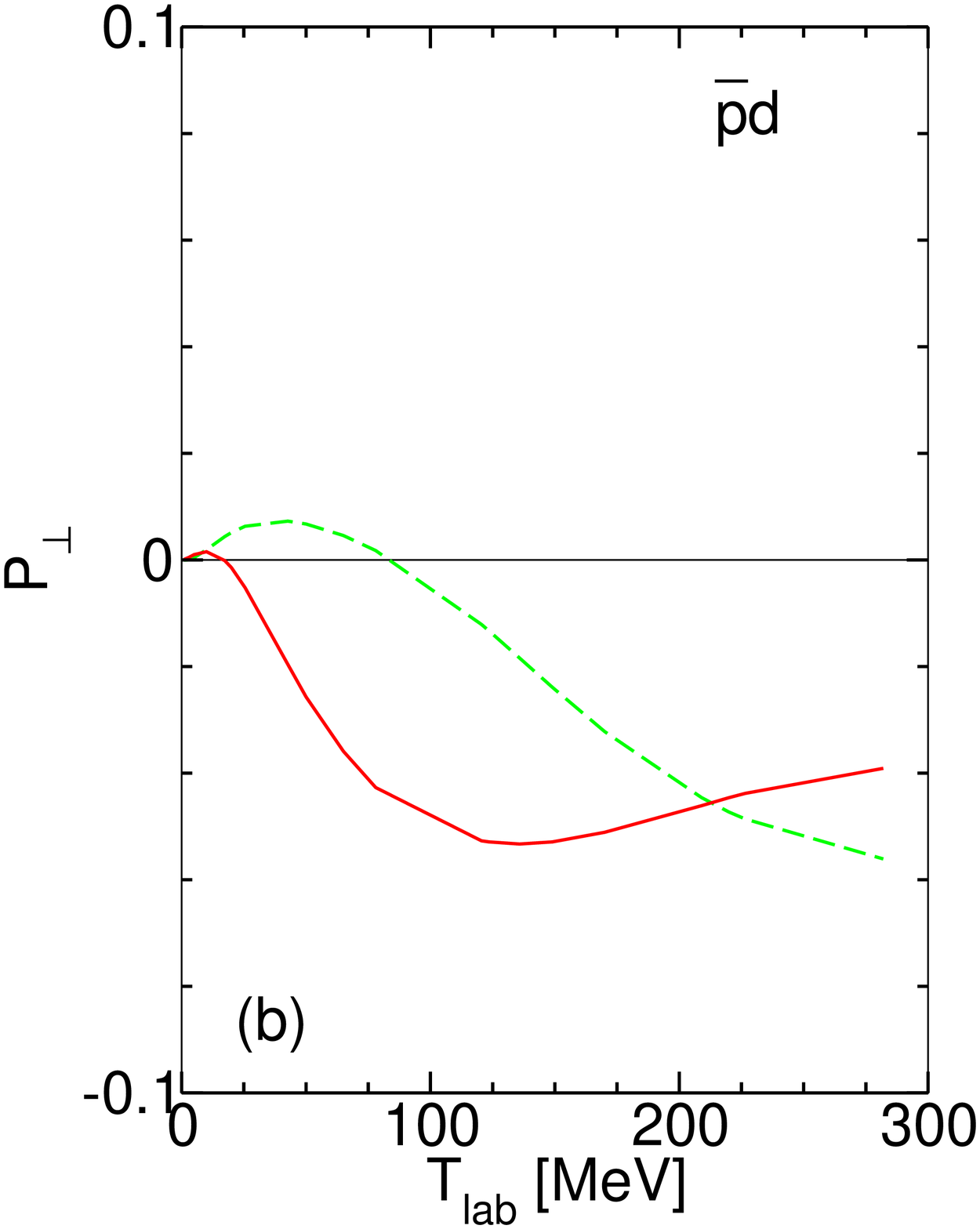}
\includegraphics[width=0.32\textwidth,angle=0]{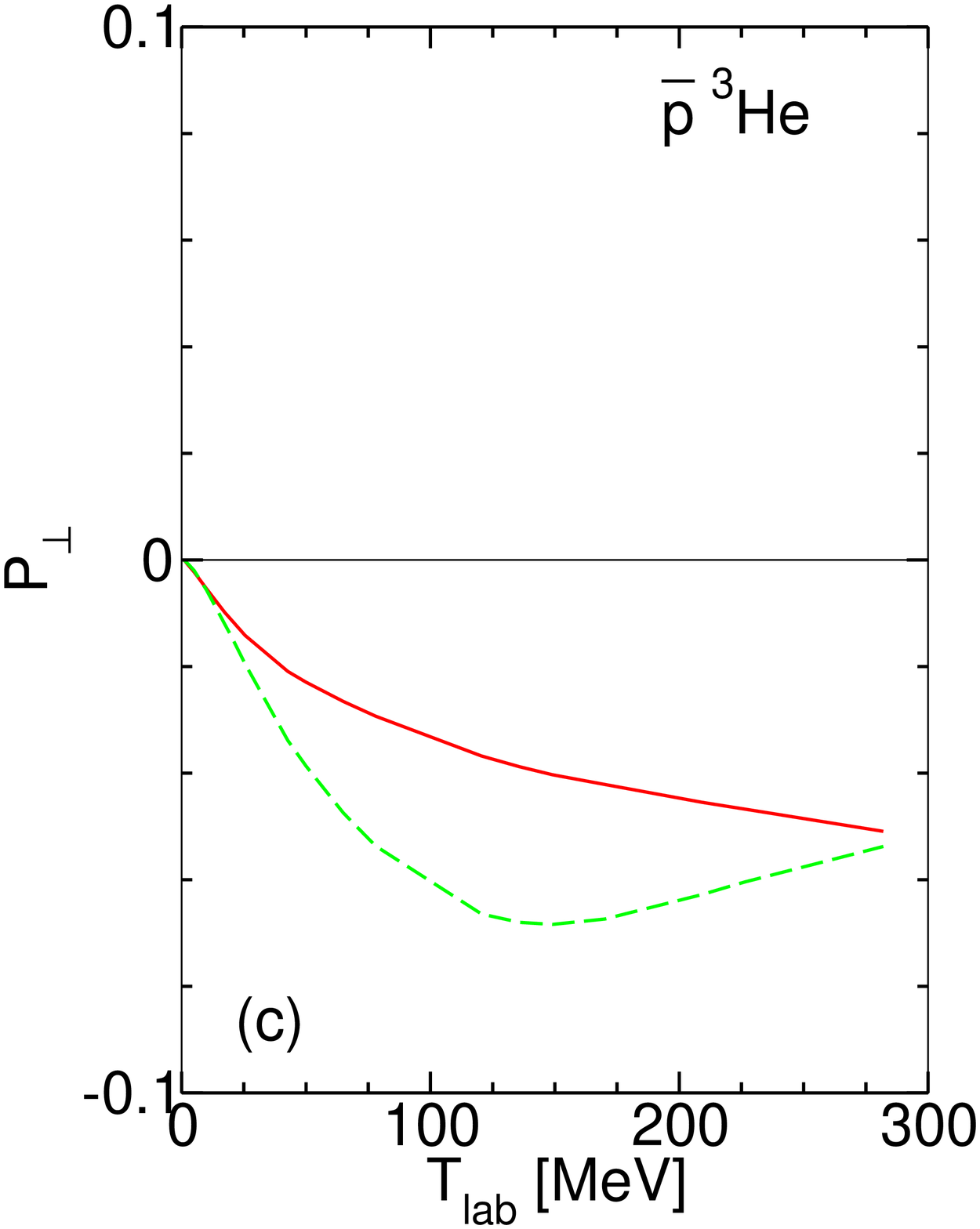}
\caption{Dependence of the
transversal polarization $P_{\perp}$ (i.e. $P_{\bar p}(t_0)$ for
${\bfg \zeta}\cdot {\bf \hat k}=0$) on the beam energy
for the target polarization $P_T=1$ in the reactions
$\bar pp$, $\bar pd$, and $\aphet$.
The results are for the models A (dashed line) and D (solid line).
The employed acceptance angle is $\theta_{acc}=10$ mrad. 
}
\label{pperp}
\end{figure}
Our results suggest that for energies 50-300 MeV
reliable predictions can be obtained within the Glauber-Sitenko 
approach for the differential cross section (Fig. \ref{pd179q2})
and also for the spin observables $A_y^d$, $A_y^{\bar p}$, $A_{xx}$, 
and $A_{yy}$ (Fig.~\ref{ay179}) for $\theta_{c.m.}$
up to $50^\circ-60^\circ$ in the $\bar p d$ system.
Obviously, within this angular region there is practically no sensitivity to the 
$\bar p N$ amplitudes in the backward hemisphere, in accordance with the
requirements of the Glauber-Sitenko approach. 
As expected, due to the influence of the deuteron elastic form factor the width 
of the corresponding bands are smaller for higher energies and larger at lower 
energies, see the corresponding results in Ref.~\cite{uzjh2012}.
According to our calculations this (reliability) region 
includes the whole diffractive peak in the differential cross section 
at forward angles, for energies from around 50 MeV upwards. 
This finding validates the application of the optical theorem for evaluating 
the total polarized cross sections based on the obtained forward $\bar p d$ 
amplitude \cite{ujh2009}.  
 
With regard to the measured differential cross section at 179 MeV,
see Fig.~\ref{pd179q2}b, our Glauber-Sitenko
calculation describes the first diffractive peak quite well - for $\bar p N$
amplitudes generated from model A as well as for those of model D. 
The first minimum in the differential cross section, 
located at $q^2\approx 0.12-0.13$ (GeV/c)$^2$
(i.e. $\theta_{c.m.}\approx 55^\circ$), and the onset of the second 
maximum is explained only by model D. The obvious strong disagreement with 
the data at larger transferred momenta, $q^2> 0.15$ (GeV/c)$^2$, 
corresponding to $\theta_{c.m.}>60^\circ$, lies already in the region where
the Glauber-Sitenko theory cannot be applicable anymore
and, therefore, no conclusions can be drawn. 
In this context let us mention that the results shown in Fig.~\ref{pd179q2}a
were obtained with the full (unmodified) $\bar pN$ amplitudes as predicted by 
the J\"ulich $\bar NN$ model D.

The results obtained for the vector analyzing powers
$A_y^{\bar p}$ and $A_y^d$ indicate a strong model dependence (Fig. \ref{ay179}).
When the spin-dependent terms of the elementary $\bar p N$ amplitude 
($B_N$, $ C_N$, $C_N'$, $G_N$, $H_N$) are excluded, then the vector analyzing 
powers $A_y^{\bar p}$ and $A_y^d$ vanish. 
In contrast, the tensor analyzing 
powers $A_{xx}$ and $A_{yy}$ are much less sensitive to the $\bar N N$ models
in question. Indeed, these observables are dominated by the spin-independent 
amplitudes, see the dash-double dotted line in Fig. \ref{ay179}c and d. 
Thus, the results obtained here for $A_{xx}$ and $A_{yy}$ 
seem to be quite robust up to scattering angles of $60^\circ-70^\circ$. 
The tensor analyzing powers $A_{xx}$ and $A_{yy}$ are reduced by one order of 
magnitude when the $D$-wave is omitted (dashed line in Fig. \ref{ay179}).
Actually, $A_{xx}$ and $A_{yy}$ practically vanish if, in addition,
the spin-dependent terms of the elementary $\bar p N$ amplitude are omitted.

To estimate the efficiency of the polarization buildup mechanism it is
instructive to calculate the polarization degree $P_{\bar p}$ at the beam life 
time $t_0$ \cite{mss}. With our definition of $\sigma_1$ and $\sigma_2$
in Eq.~(\ref{sigtot}) this quanitity is given by 
\begin{eqnarray}
P_{\bar p} (t_0)=-2P_T\frac{\sigma_1}{\sigma_0}, \ {\rm if} \ {\bfg \zeta}\cdot {\bf \hat k}=0,
\phantom{xxxx}
P_{\bar p} (t_0)=-2P_T\frac{\sigma_1+\sigma_2}{\sigma_0}, \ {\rm if} \ 
 |{\bfg \zeta}\cdot {\bf \hat k}|=1 \ ,
\label{RL}
\end{eqnarray}
where the unit vector ${\bfg \zeta}$ is directed along 
the target polarization vector $P_T$.
Results for the transversal polarization $P_{\perp}$ (${\bfg \zeta}\cdot {\bf \hat k}=0$)
are shown in Fig. \ref{pperp}.
Since the $\bar p~^3{\rm He}$ cross sections were calculated in the single-scattering
approximation \cite{ujhprm}, the results for $\bar pd$ interaction 
in Fig. \ref{pperp} are likewise presented in this approximation. 
One can see that the polarization efficiency is comparable 
in absolute value for the reactions $\bar pp$, $\bar p d$, and $\bar p\,^3{\rm He}$
\cite{jhuz2011}. 
However, since the total cross section is larger in case of $^3{\rm He}$ 
the resulting efficiency of the polarization buildup tends to be 
somewhat smaller than those for $\bar pp$ and $\bar p d$.
  
The double scattering mechanism (shadowing effects), considered for $\bar pd$ 
in \cite{uzjh2012}, decreases the absolute value of the 
polarized as well as of the unpolarized total cross sections 
and the polarization efficiency decreases too.  
Similar effects for $\bar pd$ were reported in Ref.~\cite{salnikov} using 
amplitudes from the Nijmegen $\bar pp$ partial wave analysis \cite{Timmermans}. 

\vskip 0.3cm 
\noindent {\bf Acknowledgements:}
This work was supported in part by the WTZ project no. 01DJ12057, 
the Heisenberg-Landau program, and the JINR-Kazakhstan program.

\end{document}